\documentclass[12pt]{article}

\usepackage{amssymb}
\usepackage{graphicx}		
\usepackage{dcolumn}			
\usepackage{bm}					
\usepackage{float}			
\usepackage{amsmath,amsfonts}	
\usepackage{url}					
\usepackage{setspace}
\usepackage{epstopdf}
\usepackage{caption}
\usepackage{authblk}

\newcommand{\bs}[1]{\boldsymbol{#1}}

\usepackage{natbib}
\setcitestyle{numbers}
\setcitestyle{square}
\setcitestyle{comma}

\begin{document}

\title{Study of human phonation in a full body domain}

\author{Shakti Saurabh and Daniel J Bodony}


\date{%
    Department of Aerospace Engineering, University of Illinois at Urbana-Champaign\\%
}

\markboth{Bulletin of the American Physical Society ,˜Vol
.˜60, No.˜21,˜2015}{Shell \MakeLowercase{\text
it{et al.}}: A Novel Tin Can Link}

\maketitle

\begin{abstract}
The Generation and propagation of the human voice is studied in two-dimensions using a full-body domain, using direct numerical simulation. The fluid/air in the vocal tract is modeled as a compressible and viscous fluid interacting with the non-linear, viscoelastic vocal folds (VF). The VF tissue material properties are multi-layered, with varying stiffness, and a finite-strain model is utilized and implemented in a quadratic finite element code. The fluid-solid domains are coupled through a boundary-fitted interface and utilize a Poisson equation-based mesh deformation method. The full-body domain includes the near VF region, the vocal tract, a simplified model of the soft palate and mouth, and extends out into the acoustic far-field. A new kind of inflow boundary condition based upon a quasi-one-dimensional formulation with constant sub-glottal volume velocity, which is linked to the VF movement, has been adopted. The sound pressure levels (SPL) measured are realistic and we analyze their connection to the VF dynamics and glottal and vocal tract geometries.
\end{abstract}

\section{Introduction}
\label{Intro}
With the advent of fast computational resources in the past decade and a half, the ability of computational fluid dynamics (CFD) to study phonation has gained much attention. However, the fidelity of CFD techniques are challenged by the uncertainty in determining accurate material properties and the internal layering of the VFs. This was overcome in a recent study \cite{Xue2011}, which has provided quantitative estimates to the above problems.

Initial investigations using CFD assumed rigid VFs \cite{Zhang et al. 2002a, Zhao et al. 2002,Zhao et al. 2001b} or with a prescribed motion \cite{Bodony2007,Alipour2000,Alipour2004,Larsson2009,Suh2007}. Only a small subset allowed the VF motion to be dynamically coupled with the glottal flow \cite{Luo2008,Luo2009}. More recently, a sharp-interface immersed boundary method has been developed and implemented to study both the glottal airflow and linear viscoelastic VFs \cite{Luo2008,Luo2009}. Also the first 3D continuum model of the VF was developed using finite-element method only in 2000 \cite{Alipour2000}. However, the flow model used was still Bernoulli's equations. Recently, there has been advancements in this respect, with studies being able to implement a fully coupled, 3D continuum based model of the VF and the glottal flow \cite{Rosa2003, Zheng et al. 2010, Zheng et al. 2011b, SS2014, SS2013}.

Current simulation is fully compressible, with FSI and finite deformation structure model implemented. We study the case of both adult and child phonation. There are anatomical and structural differences between an adult and a child vocal fold/tract \cite{Hirano1983}. This leads to significant differences in phonation characteristics.

\section{Methods}
\label{Num}

\subsection{Fluid Numerical Approach}
\label{FluidNum}
The fluid model uses a fully compressible formulation of the Navier-Stokes equations. The conservation of mass, momentum
and energy are described by,

\begin{eqnarray}
 \label{eq:1}
 \frac{\partial \rho}{\partial t} + \frac{\partial}{\partial x_i}(\rho u_i) &= 0, \nonumber\\
 \frac{\partial \rho u_i}{\partial t} + \frac{\partial}{\partial x_j}\left (\rho u_i u_j + p\delta_{ij} - \tau_{ij} \right) &= 0, \\
 \frac{\partial \rho E}{\partial t} + \frac{\partial}{\partial x_j}\left [\left(\rho E + p\right)u_j + q_j - u_i \tau_{ij} \right] &= 0, \nonumber
\end{eqnarray}

where $\rho$, $\rho\bs{u}$, $\rho E=p/(\gamma-1)+\frac{1}{2}\rho\bs{u}\cdot\bs{u}$,  are the conserved variables representing density,
specific momentum vector, and specific total energy, respectively. Viscous stress tensor is given by $\tau_{ij}=\mu(\partial u_i/
\partial x_j+\partial u_j/\partial x_i)+\delta_{ij}\lambda \partial u_k/\partial x_k$. The first and second viscosity coefficients, and
Kronecker delta are given by $\mu$, $\lambda$,  and $\delta_{ij}$ respectively. The thermodynamic pressure is given by
equation of state and the heat flux vector by Fourier's law. Repeated indices imply summation. 

\begin{figure*}
\begin{center}
\captionsetup{justification=centering,margin=2cm}
  \includegraphics[width = 9.5cm]{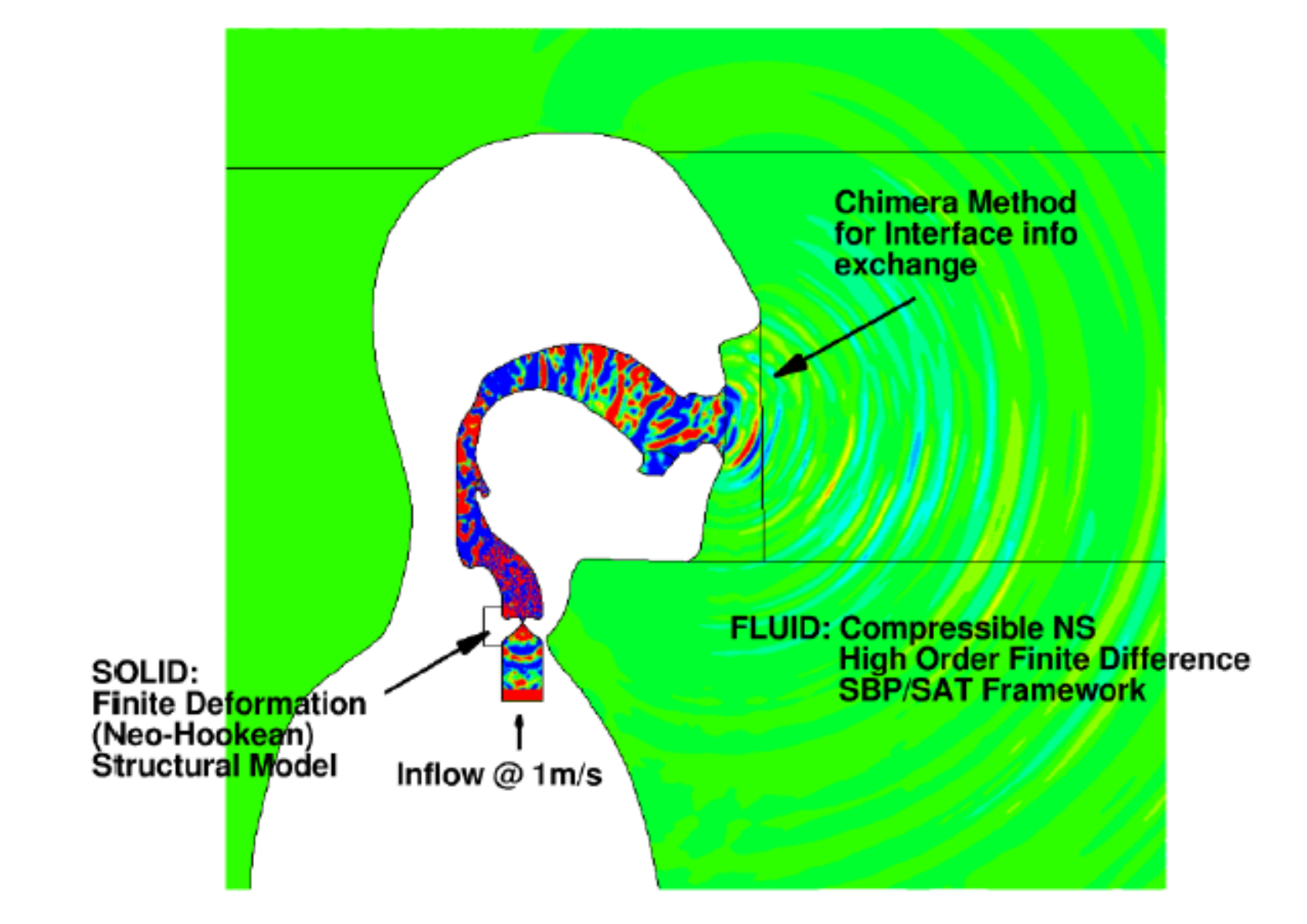}
\caption{\label{fig:FIG1}{Fluid domain with multiple blocks.}}
\end{center}
\end{figure*}

The fluid domain may consist of multiple blocks to accommodate complex geometries (Fig.~\ref{fig:FIG1}). Each block thus shares a common
overlapping interface with the adjacent block. The coupling between the blocks requires a stable interface treatment as well as
preserve conservation. This is accomplished by following a recent study \cite{NGVS2009} wherein conditions for stability
and conservation of both viscous and inviscid fluxes in a SAT framework, have been presented and verified. Some studies have applied this approach with relative success \cite{SS2013, sharan2016tim, SS2014}. 

The temporal advancement of Eq.~(\ref{eq:1}) is accomplished using an explicit four-stage Runge Kutta method
which is fourth-order accurate. The fully coupled fluid-structure phonation problem involves significant deformation of the vocal
tracts and hence necessitates the use of deforming grids for body-fitted meshes. A transfinite interpolation scheme is used to
compute the grid coordinates after deformation, that works on the principle that the arclength-normalized coordinate of each
point along its associated grid line is preserved.

\subsection{Structural Numerical Approach}
\label{SolidNum}
A non-linear, finite-strain finite element solver is used to calculate for the structural state at each point in time in the solid
domain. In the finite-strain formulation, the initial configuration $B_O$ of the body is defined by $\bs{X}$, which eventually due
to the application of a load undergoes a displacement $\bs{u}$ to reach the deformed configuration $\bs{x}=\bs{X}+\bs{u}$.

\begin{equation}
 \label{eq:6}
\nabla \cdot \bs{\sigma}+\rho\bs{b}=\rho\ddot{\bs{u}},
\end{equation}

where $\bs{\sigma}$ is the Cauchy stress tensor, $\bs{b}$ is a field of body forces, $\rho$ is the density, $
\ddot{\bs{u}}=\partial^2\bs{u}/\partial t^2$ is the acceleration, and $\nabla$ is the gradient.

\subsection{Interface treatment}
\label{Int}
There is a weak coupling between the individual fluid and structural solvers at the interface where the fluid-structure interaction 
takes place. The fluid and structural domains are solved for independently  at a given time step $t_n = n\Delta t$. At the 
interface, the spatial coupling is obtained by matching nodes between the two domains. Solid displacement and velocities are 
dirichlet quantities that are passed from the solid to the fluid, while the Neumann quantities like traction and heat flux are passed 
from the fluid to the solid, as suggested by Giles \cite{G1997}. 

\subsection{Physical model}
\label{pm}

\begin{figure*}
\begin{center}
  \includegraphics[width = 9.5cm,trim={0.2cm 0.2cm 0.2cm 0.2cm},clip]{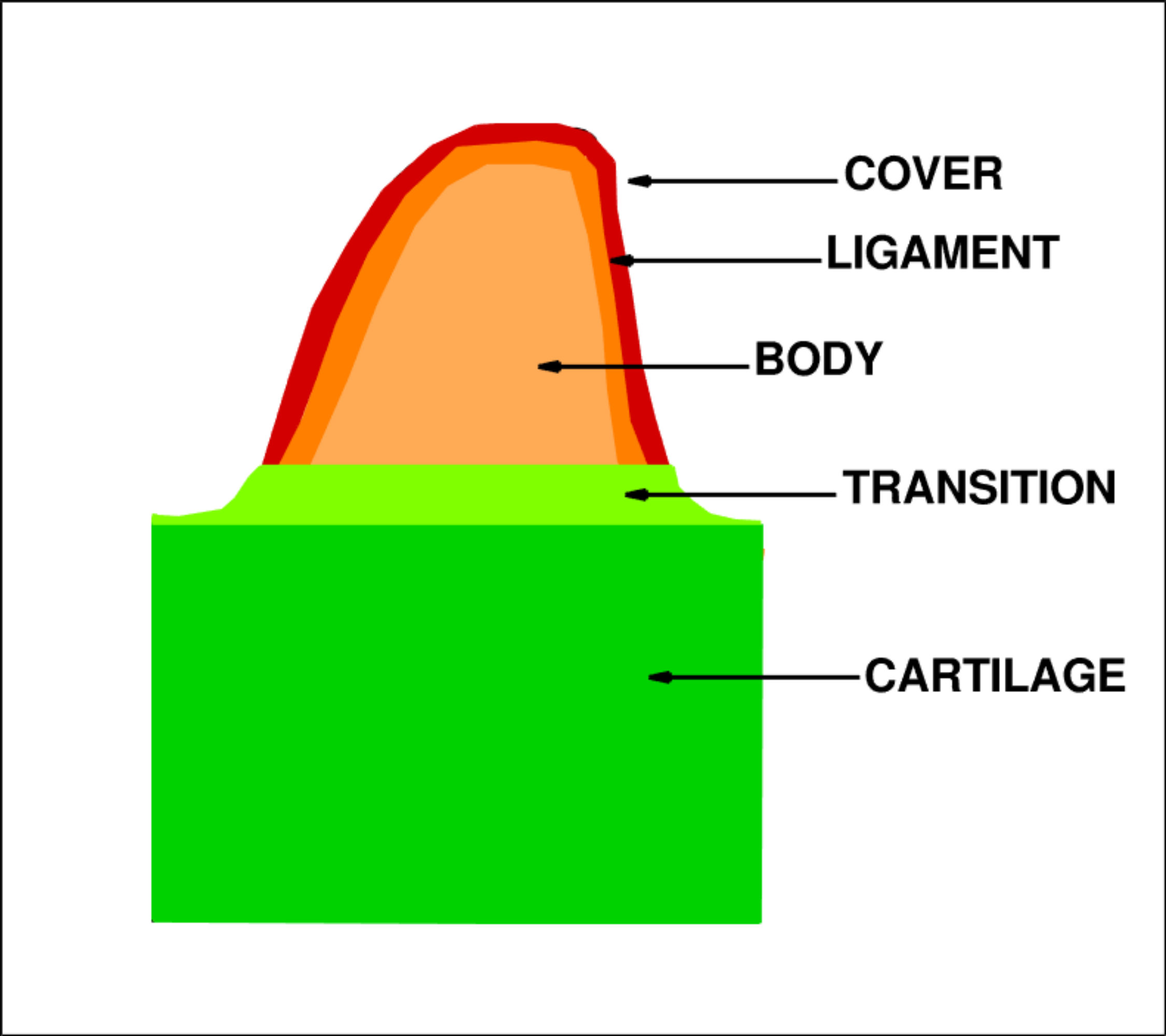}
\caption{\label{fig:FIG3}{Vocal fold model showing the different layers.}}
\end{center}
\end{figure*}

A three layer model of the VFs composed of the cover, ligament and the body as shown in Fig.~\ref{fig:FIG3} is used. The VF layers terminate into a stiff cartilage after transitioning smoothly (using a tan hyperbolic function) through a transition layer. Vocal fold material properties of an adult are listed in Table~\ref{tab:table1}. The same for a child are listed in Table~\ref{tab:table2}.

\begin{table}[htbp]
\caption{VF material properties of an adult}
\label{tab:table1}\centering%
\begin{tabular}{cccc}
 Parameter & $E_{body}$ &$E_{cover}$ &$E_{ligament}$\\
\hline
YngMod(kPa) & 50& 25 & 100 \\
Poisson Ratio & 0.27 & 0.27 &0.27 \\
\end{tabular}
\end{table}

\begin{table}[htbp]
\caption{VF material properties of a child}
\label{tab:table2}\centering%
\begin{tabular}{cccc}
 Parameter & $E_{body}$ &$E_{cover}$ &$E_{ligament}$\\
\hline
YngMod(kPa) & 50& 25 & 25 \\
Poisson Ratio & 0.27 & 0.27 &0.27 \\
\end{tabular}
\end{table}

The vocal tract experiences significant changes with age \cite{Vorperian}. First, the length of the vocal tract increases with age. Another change that is significant is the laryngeal descent \cite{Vorperian} with age. Other changes include increase of pharyngeal length with age and development of curvatures along the vocal tract past the VFs. Most structures, despite differences in growth rate, appear to follow a growth pattern or a growth curve that is similar to that of the vocal tract length. We created a 2 year old child model for our simulation, by incorporating the above mentioned details. 

\section{Results and Discussion}
\label{Results}

\begin{figure*}
\begin{center}
        \includegraphics[width=14.5cm,trim={0.2cm 0 0 0.2cm},clip]{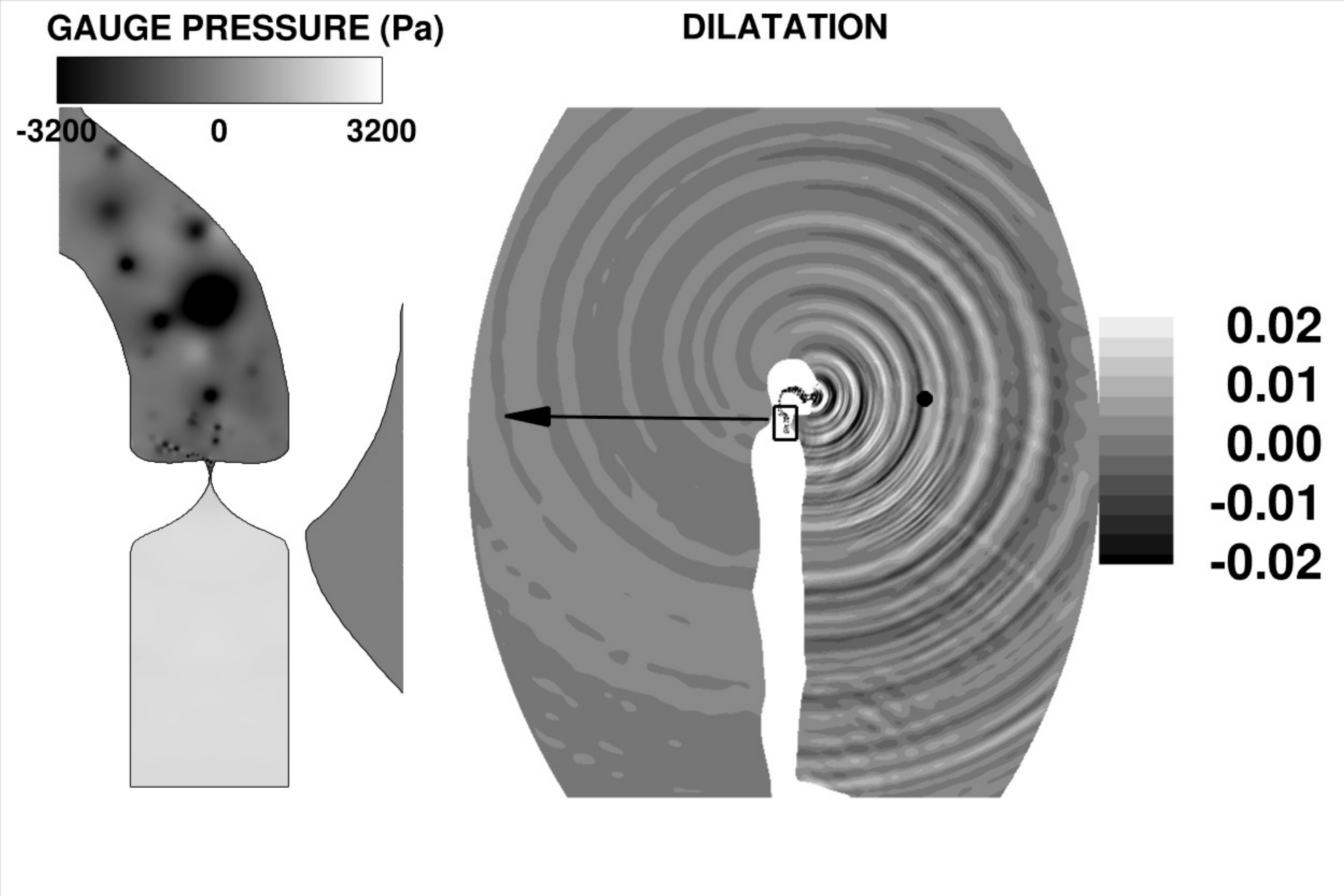}	
\caption{\label{fig:FIG4}{Gauge pressure (left) and far-field dilatation ($\triangledown.\bs{u}$) field (right) for the human adult full body simulation}}
\end{center}
\end{figure*}

\begin{figure*}
\begin{center}
        \includegraphics[width=14.5cm,trim={0.2cm 0 0 0.2cm},clip]{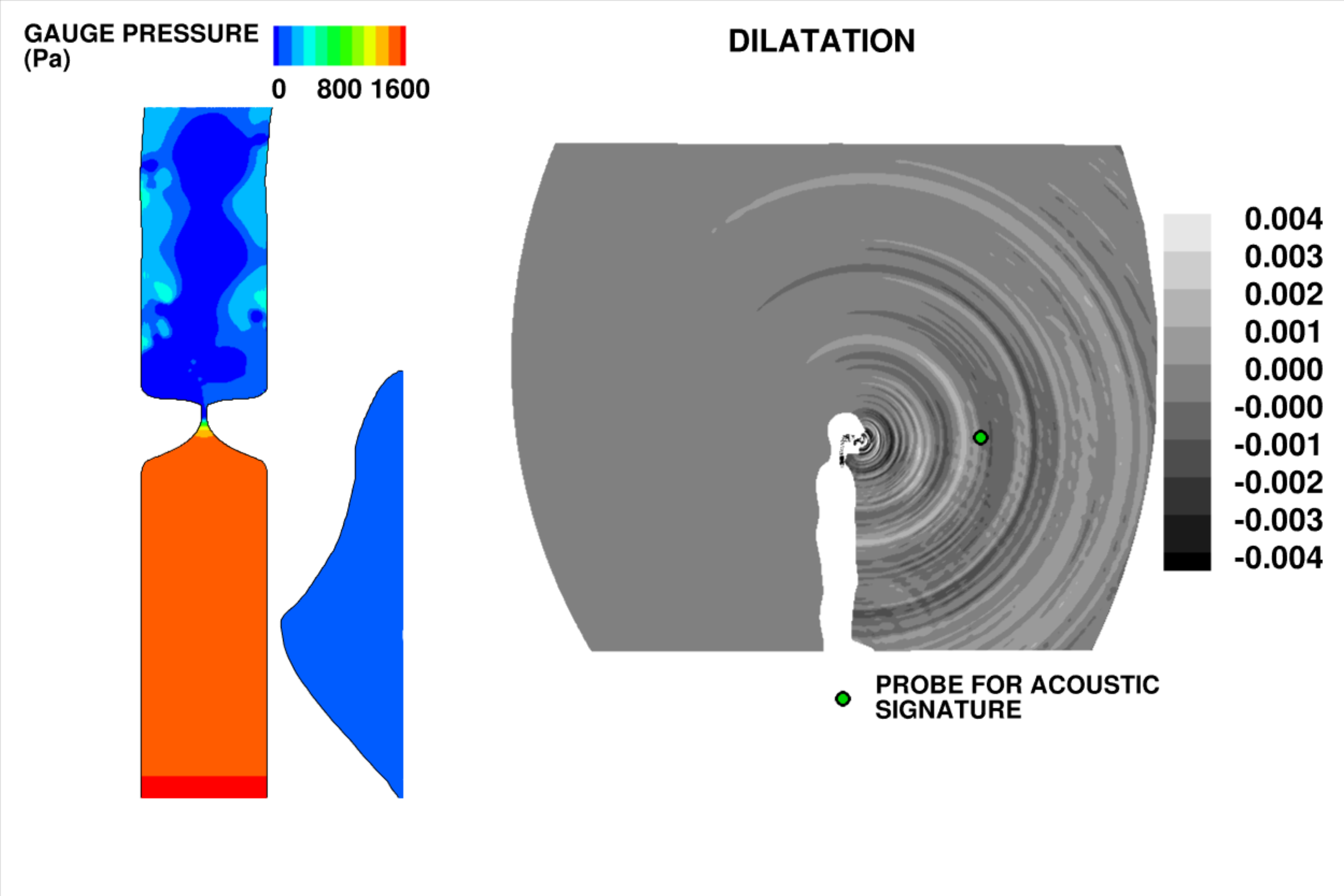}	
\caption{\label{fig:FIG5}{Gauge pressure (left) and far-field dilatation ($\triangledown.\bs{u}$) field (right) for the human child full body simulation}}
\end{center}
\end{figure*}

The two major parameters that are significant for voice quality are loudness and pitch. The former is characterized by the sound 
pressure level (OASPL) and the latter by the frequency content of the acoustic field. Fig.~\ref{fig:FIG4} shows the oscillatory 
behavior of the acoustic waves (dilatation) exiting the mouth of an adult which are induced as a result of the VF motion and the interaction 
of the glottal jet with the vocal tract walls. This data can be used to study acoustic characteristics of the flow which is computed 
directly without resorting to any acoustic analogy. The same for a child can be seen in Fig.~\ref{fig:FIG5}. The OASPL and fundamental frequency for both the adult and child are listed in Table~\ref{tab:table3}. As can be seen the major difference arises in the frequency of emitted sound of a child. This is consistent with data from literature as well.

\begin{table}[htbp]
\caption{Comparison of OASPL and  fundamental frequency between an adult and a child phonation }
\label{tab:table3}\centering%
\begin{tabular}{cccc}
 Case & $OASPL (dB)$ &$F0 (Hz)$\\
\hline
Adult & 69.12& 240 \\
Child & 71.8 & 375 \\
\end{tabular}
\end{table}

Following the simulation of the adult and child phonation, the effect of changing the material properties of the VF layers on phonation characteristics like sound intensity and pitch was analyzed. Three different cases for this analysis was undertaken. Table~\ref{tab:table4}  provides the parameters for each of these cases and also comparing to the already completed adult simulation. The cover layer elasticity was varied by $\mp 20 \%$ for the first two cases and the ligament layer was made considerably softer for the third case, while the other properties were kept unchanged. Varying one property at a time, will give an insight on the effect of each of these variations independently. All other simulation parameters were unchanged from the adult case.

\begin{figure*}
\begin{center}
         $(a)_{\includegraphics[width=8.0cm]{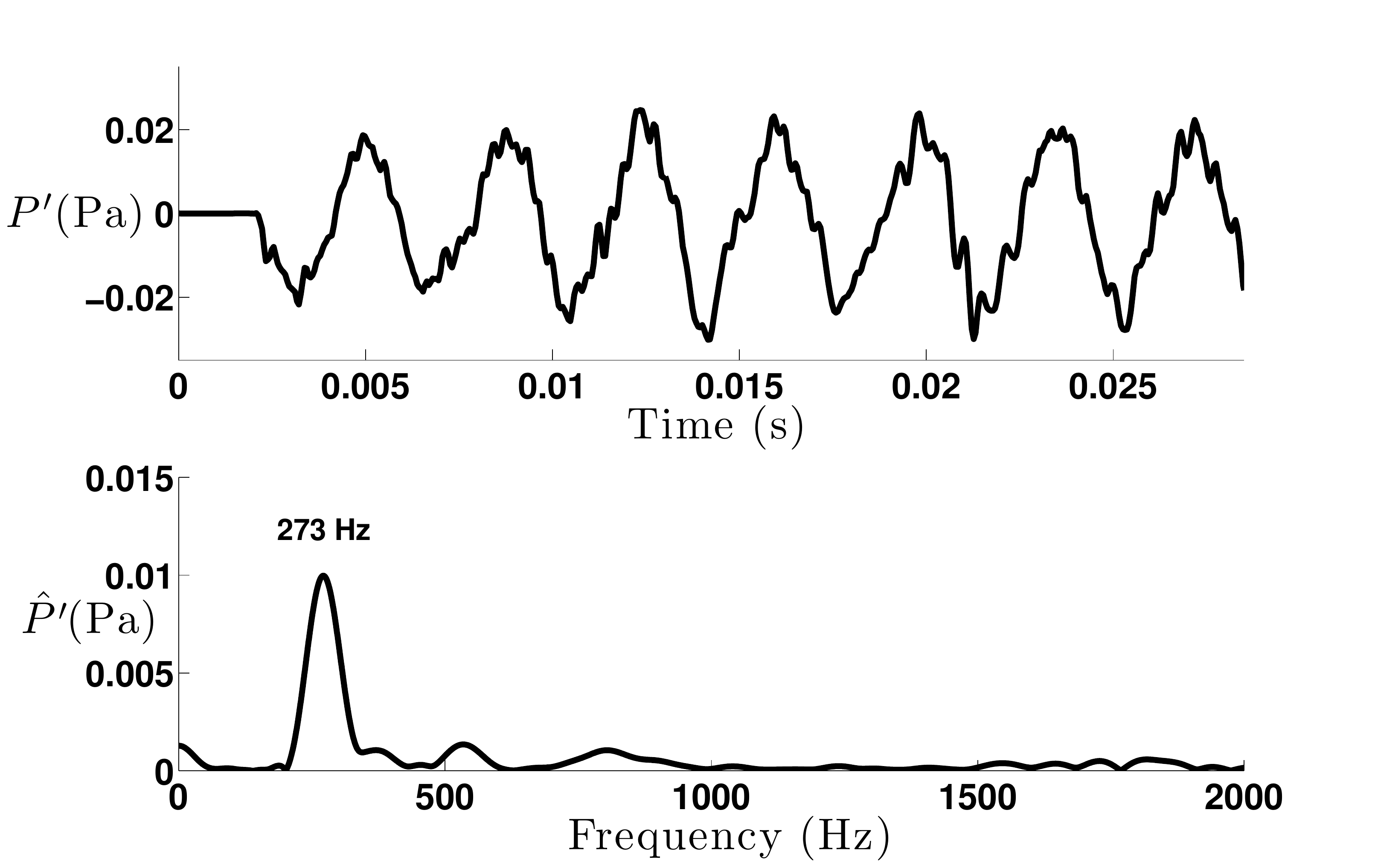}}$
	$(b)_{\includegraphics[width=8.0cm]{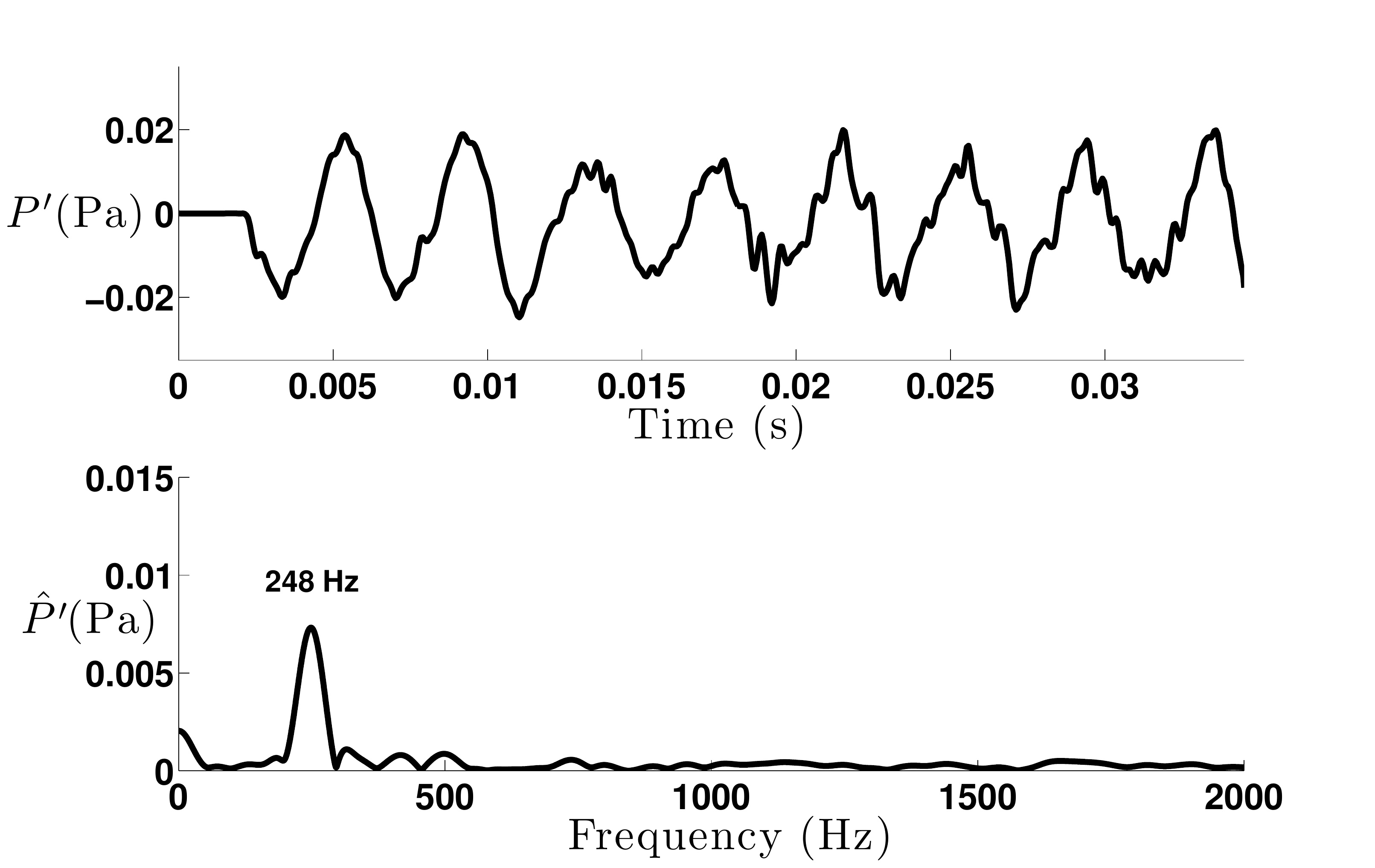}}$\\	
	$(c)_{\includegraphics[width=8.0cm]{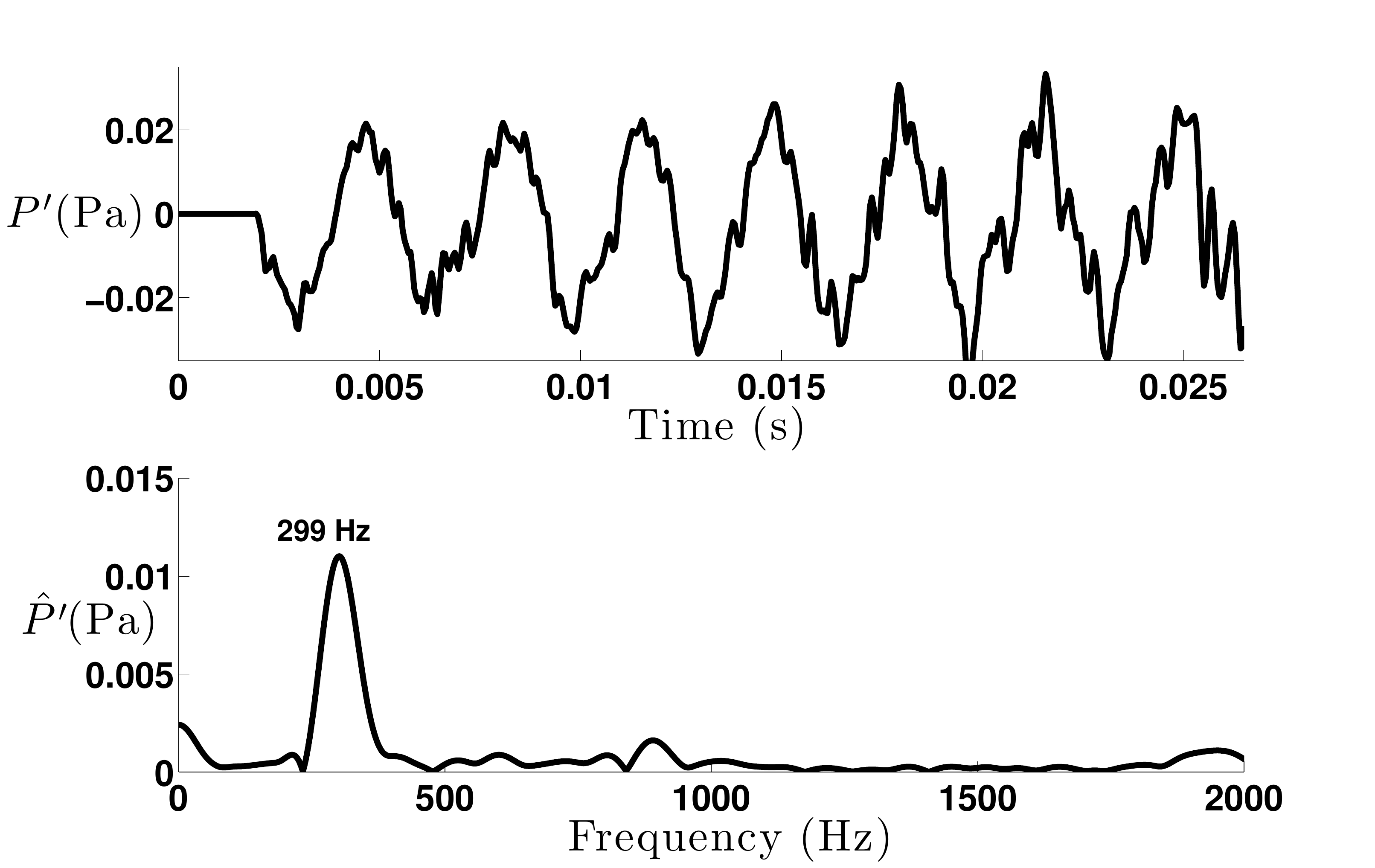} }$
	$(d)_{\includegraphics[width=8.0cm]{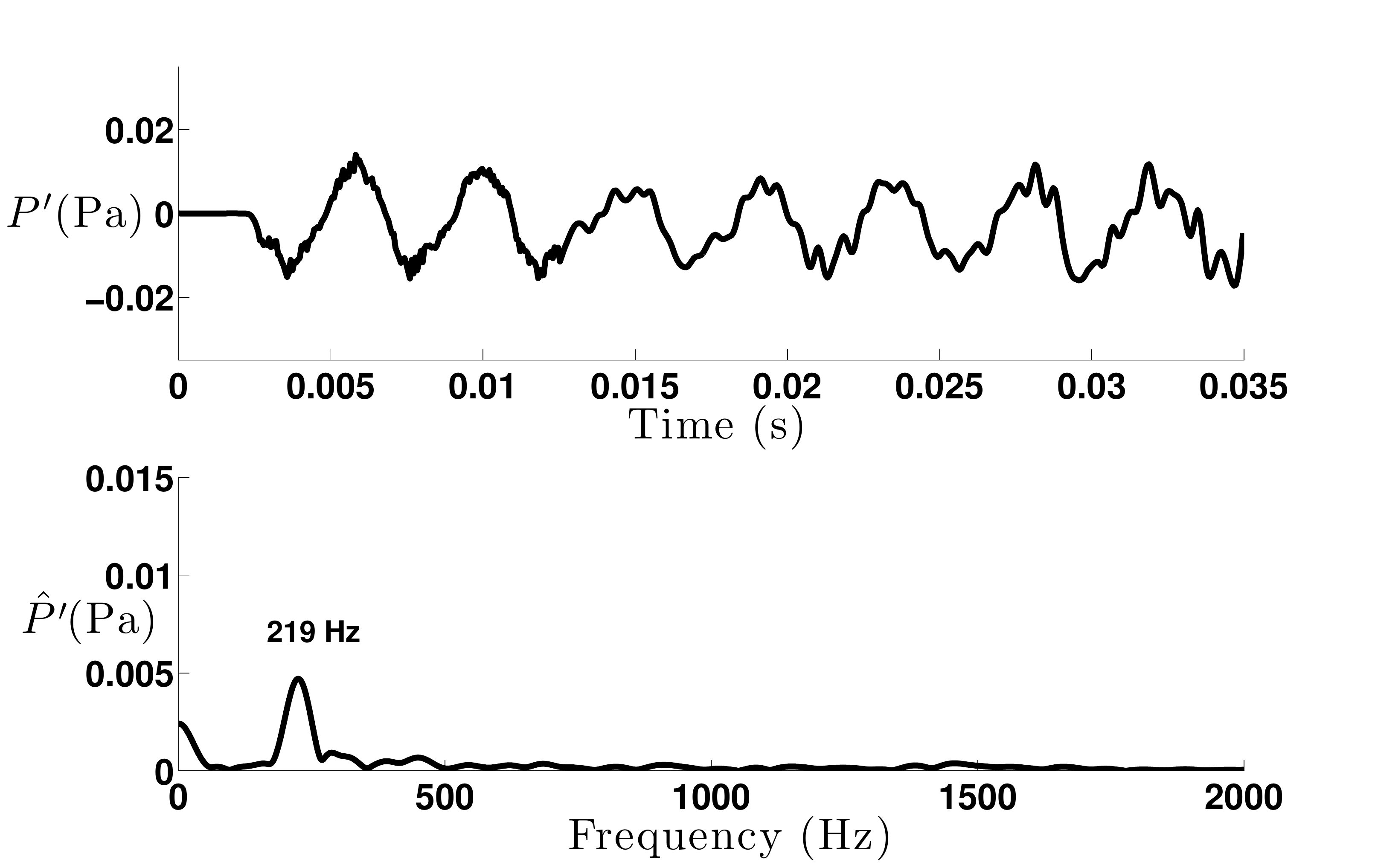} }$\\
\caption{ \label{fig:FIG6} {Far-field pressure fluctuation versus time for VF material property sensitivity study for (a) Adult, (b) Case 1, (c) Case 2 and (d) Case 3.}}
\end{center}
\end{figure*}

The pressure fluctuation are plotted for this study in Fig.~\ref{fig:FIG6}. The values of the fundamental frequencies F$0$, also shows significant variations from case to case. Our understanding points to observing higher frequencies when the VFs are stiffened and lower when they are made softer. Observations from simulations follow this trend as well. When compared to the adult case, stiffening the VFs for case 2, increased the fundamental frequency while making it softer for case 1 and 3 decreased it. Table~\ref{tab:table5} also provides the OASPL obtained from the corresponding pressure fluctuations for the cases studied here. Thus the significant differences in far-field quantities indicate the relevance of change in VF stiffness on the calculated acoustic field. 

\begin{table}[htbp]
\caption{\label{tab:table4} VF Material Properties (in kPa) for 3 cases with perturbed properties}
\begin{center}
\begin{tabular}{| c | c | c | c |}
\hline
Case & $E_{Cover}$ & $E_{Ligament}$ & $E_{Body}$ \\ \hline
Adult & 25 & 100 & 50 \\ 
Case 1 & 20 & 100 & 50 \\ 
Case 2 & 30 & 100 & 50\\ 
Case 3 & 25 & 10 & 50\\ \hline
\end{tabular}
\end{center}
\end{table}

\begin{table}[htbp]
\caption{\label{tab:table5} OASPL for the sensitivity study}
\begin{center}
\begin{tabular}{|c|c|c|}
\hline
Case & $SPL (dB)$\\ \hline
Adult & 69.12 \\ 
Case 1 & 69.85 \\ 
Case 2 & 68.47 \\ 
Case 3 & 70.09 \\ \hline 
\end{tabular}
\end{center}
\end{table}

\section{Conclusion}
\label{Conclusion}

Full body simulation provides an extensive outlook to phonation due to anatomically realistic vocal tract geometry and presence of head and body. Quasi-1D dynamic boundary condition may provide an answer to the issue of sub-glottal boundary condition setup. Higher fundamental frequency and sound pressure levels were observed (compared to adult) from child simulation. Sensitivity analysis can give further insight upon the dependence of phonation characteristic on material parameters etc.

\section*{Acknowledgement}
\label{Ack}
This work was supported by the National Science Foundation (CAREER award number 1150439, Dr.~D.~Papavassiliou as the 
technical monitor) Computational support for this project has been provided by NSF XSEDE resources, award number TG-
CTS090004.


\newpage
\newpage

\end{document}